\documentclass[aps,prd,letterpaper,twocolumn,preprintnumbers,nofootinbib,superscriptaddress]{revtex4-2}
\usepackage[a4paper, hdivide={1.91cm,,1.165cm}, vdivide={1.83cm,,2.9cm}]{geometry}
\usepackage{subfigure}
\usepackage{slashed}
\usepackage{amsmath,amssymb}
\usepackage{graphicx}
\usepackage{units}
\usepackage{bbold}
\usepackage{xcolor}
\usepackage{dsfont}
\usepackage[hyperfootnotes=false,colorlinks,citecolor=blue]{hyperref}
\usepackage{comment}
\usepackage[normalem]{ulem} 
\usepackage{comment}
\usepackage{autobreak}

\newcommand{\beq}{\begin{equation}}
\newcommand{\eeq}{\end{equation}}
\newcommand{\bea}{\begin{eqnarray}}
\newcommand{\ena}{\end{eqnarray}}

\def \epsilon {\varepsilon}

\newcommand{\hc}{\ensuremath{\text{H.c.}}}

\newcommand{\change}[1]{#1}

\allowdisplaybreaks

\begin{document}
\preprint{CETUP-2023-025}
\title{Decaying scalar dark matter in the minimal left-right symmetric model}

\author{P. S. Bhupal Dev}
\email[E-mail: ]{bdev@wustl.edu}
\thanks{ORCID: \href{https://orcid.org/0000-0003-4655-2866}{0000-0003-4655-2866}.}
\affiliation{Department of Physics and McDonnell Center for the Space Sciences, Washington University, St.~Louis, Missouri 63130, USA}

\author{Julian Heeck}
\email[E-mail: ]{heeck@virginia.edu}
\thanks{ORCID: \href{https://orcid.org/0000-0003-2653-5962}{0000-0003-2653-5962}.}
\affiliation{Department of Physics, University of Virginia,
Charlottesville, Virginia 22904, USA}

\author{Anil Thapa}
\email[E-mail: ]{a.thapa@colostate.edu}
\thanks{ORCID: \href{https://orcid.org/0000-0003-4471-2336}{0000-0003-4471-2336}.}
\affiliation{Department of Physics, University of Virginia,
Charlottesville, Virginia 22904, USA}
\affiliation{Physics Department, Colorado State University, Fort Collins, Colorado 80523, USA}

\begin{abstract}
In the minimal left-right symmetric theory, the dark matter candidate is usually ascribed to the lightest right-handed neutrino. Here we present an alternative decaying dark matter candidate in this model in terms of the lightest neutral scalar from the $SU(2)_R$-triplet field. This setup requires a vast hierarchy between the scalar mass and the left-right symmetry breaking scale, which renders the scalar dark matter sufficiently stable on cosmological time scales. The stability of the dark matter imposes  constraints on the right-handed neutrino mass, which has consequences for the neutrino mass generation, as well as for leptogenesis. Although somewhat fine-tuned, it provides a very economical scenario wherein the minimal left-right model can simultaneously explain dark matter, neutrino masses, and the matter-antimatter asymmetry of the Universe.
\end{abstract}

\maketitle

\section{Introduction}

The observation of non-zero neutrino masses through neutrino oscillations is the strongest proof that the otherwise wildly successful Standard Model (SM) of particle physics  is incomplete. Ad-hoc SM extensions to accommodate neutrino masses provide interesting toy models, but ultimately a more complete SM successor that naturally generates neutrino masses is desirable. One popular model of this kind is the left-right (LR) symmetric  model~\cite{Pati:1974yy, Mohapatra:1974gc,Mohapatra:1974hk, Senjanovic:1975rk,Senjanovic:1978ev}, which extends the SM gauge group $SU(3)_c\times SU(2)_L \times U(1)_{Y}$ to the manifestly parity-symmetric $SU(3)_c\times SU(2)_L \times SU(2)_R\times U(1)_{B-L}$ and \textit{requires} right-handed (RH) neutrinos for anomaly cancellation. Neutrino masses are thus theoretically inevitable in LR models. If the LR gauge group is broken by scalar triplets $\Delta_{L,R}$ rather than doublets, lepton number is spontaneously broken and one obtains Majorana neutrino masses via both the type-I~\cite{Minkowski:1977sc, Mohapatra:1979ia, Yanagida:1979as, Gell-Mann:1979vob} and type-II~\cite{Magg:1980ut, Schechter:1980gr, Mohapatra:1980yp, Lazarides:1980nt} seesaw mechanisms, shedding light not just on the existence, but also on the \textit{smallness} of neutrino masses.

LR models also have useful ingredients to explain the second biggest shortcoming of the SM, namely, the existence of dark matter (DM).
For one, the LR breaking via triplets allows to elegantly incorporate weakly-interacting massive particles, stabilized by the leftover $U(1)_{B-L}$ subgroup $\mathbb{Z}_2$~\cite{Heeck:2015qra,Garcia-Cely:2015quu,Ferrari:2018rey}. Even without introducing additional particles, LR models can accommodate \emph{unstable} DM candidates, the almost exclusively discussed example being the lightest RH neutrino~\cite{Bezrukov:2009th,Nemevsek:2012cd,Nemevsek:2022anh,Nemevsek:2023yjl}. Aside from the additional gauge interactions, this case is similar to the popular ``keV sterile neutrino'' DM candidate~\cite{Drewes:2016upu,Boyarsky:2018tvu}, which eventually decays into an active neutrino and a photon~\cite{Pal:1981rm}, and thus can be constrained by non-observation of nearly monochromatic X-ray signatures~\cite{Dessert:2018qih, Foster:2021ngm, Krivonos:2024yvm}.

As we discuss here, the simplest LR model actually features another potential DM candidate besides the sterile neutrino, namely, the Higgs boson associated with the $SU(2)_R\times U(1)_{B-L}$ breaking, i.e.~the real part of the neutral triplet component, $\Re(\Delta_R^0)$. It is massive, electrically neutral, and can be sufficiently decoupled from the SM to be long-lived on cosmological time scales.
The possibility of long-lived $\Re(\Delta_R^0)$ being a DM candidate was briefly mentioned before in Refs.~\cite{Nemevsek:2012cd,Nemevsek:2023yjl}, but without a detailed discussion of its phenomenology.
We fill this gap in the literature and provide a first detailed study of this kind of DM and the restrictions it imposes on the LR parameter space. 
We establish $\Re(\Delta_R^0)$ as a viable decaying DM candidate in the keV to multi-MeV mass range, eventually decaying to two photons. While severe fine-tuning is required to achieve the large hierarchy between the DM mass and the LR breaking scale, we show that it restricts the RH neutrino mass scale, thereby influencing the neutrino mass generation via seesaw mechanism, as well as the generation of the observed baryon asymmetry of the Universe via leptogenesis~\cite{Fukugita:1986hr}. In this way, this economical scenario of the minimal LR model can simultaneously explain all three major pieces of empirical evidence for beyond SM physics, i.e., DM, neutrino masses, and the baryon asymmetry of our Universe. 
Moreover, we conjecture that the same DM candidate can be realized in LR extensions such as the Pati-Salam model~\cite{Pati:1974yy} or the $SO(10)$ grand unified theory~\cite{Fritzsch:1974nn}.

\section{Review of the Minimal LR Model}

The LR model is based on the gauge group $SU(3)_c \times SU(2)_L \times SU(2)_R \times U(1)_{B-L}$~\cite{Pati:1974yy, Mohapatra:1974gc,Mohapatra:1974hk, Senjanovic:1975rk,Senjanovic:1978ev}. The fermion fields transform as either left or right-handed $SU(2)$ doublets:
\begin{equation}
    Q_{L,R} = \begin{pmatrix}
        u \\
        d
    \end{pmatrix}_{L,R}  , \hspace{10mm}
    %%%
    \psi_{L,R} = \begin{pmatrix}
        \nu \\
        e
    \end{pmatrix}_{L,R}  .
\end{equation}
It naturally incorporates the RH neutrino $\nu_R$ within an RH lepton doublet, thus enabling neutrino mass generation through the seesaw mechanism~\cite{Mohapatra:1979ia, Mohapatra:1980yp}.

The Higgs sector comprises of the triplets $\Delta_L(3, 1, 2)$, $ \Delta_R(1, 3, 2)$ and the bidoublet $\Phi(2, 2, 0)$ under $SU(2)_L \times SU(2)_R \times U(1)_{B-L}$:
\begin{align}
\Phi = \begin{pmatrix}
\phi_1^0 & \phi_1^+ \\
\phi_2^- & \phi_2^0
\end{pmatrix}, \ \ 
\Delta_{L,R} = \begin{pmatrix}
\delta^+/\sqrt{2} & \delta^{++} \\
\delta^0 & -\delta^+/\sqrt{2}
\end{pmatrix}_{L,R}.
\end{align}
After the neutral component of $\Delta_R$ develops a vacuum expectation value (VEV) $\langle\delta_R^0\rangle \linebreak[1] = \linebreak[1] \smash{v_R/\sqrt{2}}$, the $SU(2)_R\times U(1)_{B-L}$ symmetry is broken to $ U(1)_{Y}$, giving masses to the $W_R^\pm$ and $Z_R$ gauge bosons. The VEVs $\langle\phi_{1,2}^0\rangle = \smash{\kappa_{1,2}/\sqrt{2}}$ break the remaining $SU(2)_L\times U(1)_{Y}$ down to the usual $U(1)_\text{\rm em}$, thereby setting the mass scale for $SU(2)_L$ gauge bosons $W_L^\pm$ and $Z_L$. The VEVs must obey the hierarchy $v_R \linebreak[1] \gg \linebreak[1] \kappa_{1,2} \gg v_L$, with 
\begin{equation}
   \kappa_+^2 \equiv  \kappa_1^2 + \kappa_2^2 \simeq (\unit[246]{GeV})^2 
\end{equation}
defining the electroweak scale, to satisfy constraints from low-energy weak interactions. Here, $\langle \delta_L^0 \rangle = v_L/\sqrt{2}$ is the triplet VEV that cannot exceed $\mathcal{O}(\unit{GeV})$ due to electroweak precision measurements, and in particular, from the $\rho$-parameter constraint~\cite{Hollik:1986gg, 
Heeck:2022fvl,ParticleDataGroup:2024cfk}.\footnote{Here we do not consider the CDF anomalous $W$-mass measurement~\cite{CDF:2022hxs}, which is in disagreement with the recent CMS measurement~\cite{CMS:2024lrd}.} Similarly, $v_R\gtrsim \unit[20]{TeV}$ in order to satisfy the current LHC constraints on the heavy gauge boson masses~\cite{CMS:2021dzb, ATLAS:2023cjo}. 

For the theory to be LR symmetric, the Lagrangian must be invariant under parity (discrete LR symmetry)\footnote{It is also possible to define a discrete charge conjugation symmetry $C$ within the model under which the scalar and fermion fields transform into their conjugate fields: $Q_{L,R} \leftrightarrow (Q^c)_{R,L}$, $\Psi_{L,R} \leftrightarrow (\Psi^c)_{R,L}$, $\Phi \leftrightarrow \Phi^*$, $\Delta_{L,R} \to (\Delta^c)_{R,L}$. If this symmetry is adopted, the Yukawa couplings $h_q$, $\tilde{h}_q$, $h_\ell$, and $\tilde{h}_\ell$ would be complex symmetric matrices, rather than hermitian matrices, with little impact on our discussion.}:
\begin{equation}
    Q_L \leftrightarrow Q_R, \hspace{3mm} \Psi_L \leftrightarrow \Psi_R, \hspace{3mm} \Delta_L \leftrightarrow \Delta_R, \hspace{3mm} \Phi \leftrightarrow \Phi^\dagger \, . 
\end{equation}
The relevant terms in the Lagrangian are
\begin{align}
\mathcal{L}_Y =& \overline{Q}_L (h_q \Phi + \widetilde{h}_q \widetilde{\Phi}) Q_R + \overline{\psi}_L (h_\ell \Phi + \widetilde{h}_\ell \widetilde{\Phi}) \psi_R \notag \\
&+ f\ [\psi_{R}^T C i\sigma_2 \Delta_{R} \psi_{R} +  \psi_{L}^T C i\sigma_2 \Delta_{L} \psi_{L}] + \text{h.c.}, \label{eq:LagLR}
\end{align}
where  $\widetilde{\Phi} = \sigma_2 \Phi^\ast \sigma_2$ with the second Pauli matrix $\sigma_2$. The mass matrices for charged leptons ($M_\ell$) and Dirac neutrinos ($M_{\nu_D}$), as well as the $6\times 6$ mass matrix for the $\nu$-$N$ sector (with $\nu \equiv \nu_L$ and $N\equiv C (\overline{\nu}_R)^T$), are 
\begin{align}
M_{\ell, \nu_D} = \frac{h_\ell \kappa_{2, 1} + \widetilde{h}_\ell \kappa_{1, 2}}{\sqrt{2}}\,, \ \  M_\nu = \begin{pmatrix}
\sqrt{2} f v_L & M_{\nu_D} \\
M_{\nu_D}^T & \sqrt{2} f v_R
\end{pmatrix}  .
\end{align}
In the limit $M_{\nu_D} \ll f v_R$, the states $\nu$ and $N$ will approximately be mass eigenstates with mass matrices:
\begin{align}
    m_N \simeq \sqrt{2} f v_R, \hspace{5mm} m_\nu \simeq \sqrt{2}  f v_L - M_{\nu_D}m_N^{-1}M_{\nu_D}^T \, .
\end{align}
Here, $m_\nu$ is the Majorana mass matrix for the light, mostly left-handed neutrinos, containing contributions from both type-II (the $v_L$ term) and type-I (the $m_N^{-1}$ term) seesaw.

The Higgs sector of the model contains 20 degrees of freedom: 8 within neutral scalars, 8 within  singly-charged scalars, and 4 within  doubly-charged scalars. After LR symmetry is spontaneously broken down to $SU(3)_c \times U(1)_{\rm em}$, 6 Goldstone modes are eaten-up by the  $W_{L,R}^\pm$ and  $Z_{L,R}$ gauge bosons. We analyze the full scalar potential in appendix~\ref{sec:AppA} to identify all the Goldstone modes and give the masses of the physical scalar fields. The upshot is that one of the neutral scalars can be sufficiently light and long-lived, thus opening the possibility to make it a decaying DM candidate.

\section{The $SU(2)_R$-triplet Higgs as dark matter}

We wish to identify the neutral scalar component of the RH triplet Higgs $\delta_R^{0} \equiv \chi$ as a prospective DM candidate. Given its involvement in spontaneous symmetry breaking and the Yukawa couplings in Eq.~\eqref{eq:LagLR}, $\chi$ can decay into SM particles. Requiring the DM lifetime to be longer than the age of the Universe  is easiest to achieve if $\chi$ is very light.
For now, we stay agnostic about the DM production mechanism, some choices being discussed below. For simplicity we do, however, assume a (semi-)thermal mechanism rather than condensation or misalignment~\cite{Baer:2014eja}.  As a result, the DM mass should exceed $\mathcal{O}(\unit[100]{eV})$, likely even several keV, to satisfy structure-formation constraints.

We  define the mixing angles of $\delta_R^0 \equiv \chi$ with $h_1^{0r}$ and $h_2^{0r}$ as $\theta_1$ and $\theta_2$, respectively (cf. Eqs.~\eqref{eq:mixingmatix} and ~\eqref{eq:sint2}). The fields $(h_1^{0r}$, $h_2^{0r}$) in the rotated basis are defined in Eq.~\eqref{eq:rotatedbasis} such that $h_1^{0r}$ is approximately identified as the SM-like Higgs boson $h$ and $h_2^{0r}$ as the heavy CP even scalar $H$. In the limit $2\rho_1v_R^2 \ll m_h^2, m_H^2$, the mass of  $\chi$ can be written as
\begin{align}
    m_\chi^2 \simeq 2 \rho_1 v_R^2 - \theta_1^2 m_h^2 - \theta_2^2 m_H^2\, ,
    \label{eq:DM}
\end{align}
where $\rho_1$ is one of $\chi$'s quartic couplings [cf.~Eq.~\eqref{eq:pot}].
Generically, $\chi$ will thus have a mass around $v_R$, so bringing it down to the electroweak scale requires considerable fine-tuning in the form of tiny parameters or cancellations. This is the price we pay for this DM candidate.
Rather than canceling these terms against each other, we  assume all of them to be minuscule, so $\rho_1,\theta_{1,2}\lll 1$. For concreteness, we  effectively set $\theta_{1,2} = 0$ in the following. 

\begin{figure}
    \centering
    \includegraphics[scale=0.21]{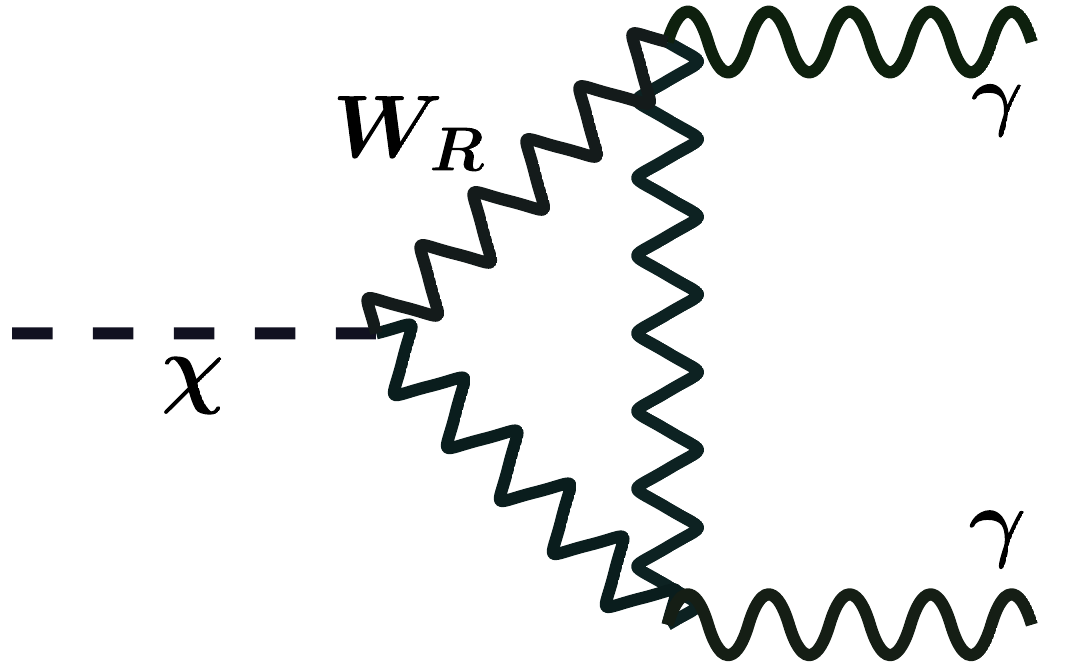} \hspace{2mm}
     \includegraphics[scale=0.21]{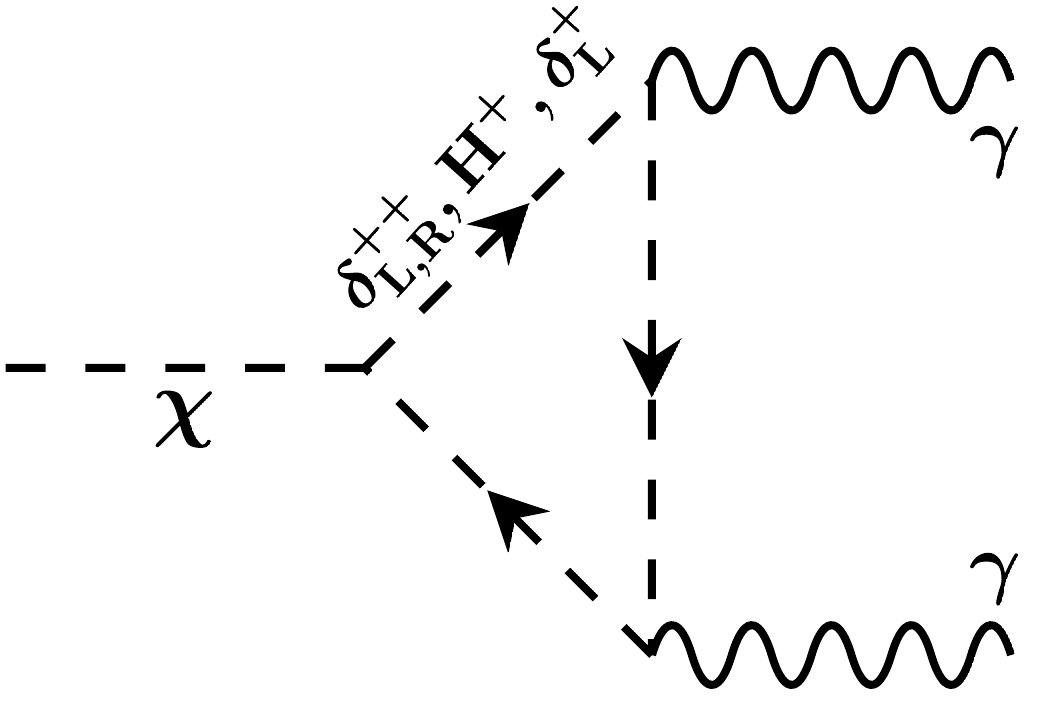}
    \caption{Loop-induced decay of scalar DM into photons through gauge bosons $W_{R}^\pm$ and  charged scalars $\delta_{L,R}^{++}$, $\delta_L^+$, and $H^+$ in the LR model. 
    }
    \label{fig:decaygg}
\end{figure}

Even for vanishing mixing angles of $\chi$ with the other two neutral scalars, it can  decay into a pair of photons at the one-loop level via virtual charged scalars $H^+$, $\delta_L^+$, and $\delta_{L,R}^{++}$, as well as the gauge boson $W_R^{+}$, as shown in Fig.~\ref{fig:decaygg}.
In the low-mass limit  $m_\chi \ll v_R$ of interest to us, the decay width reads (see Ref.~\cite{Dev:2017dui} \change{and appendix~\ref{app:DMdecay}})
\begin{equation}
    \Gamma_{\chi \to \gamma \gamma}  \simeq \frac{\alpha^2 m_\chi^3}{1028 \pi^3} \frac{4}{v_R^2} \left|  \frac{1}{3} +\frac{1}{3} + \frac{4 }{3}+ \frac{4 }{3} - 7  \right|^2 ,
\end{equation}
where $\alpha\equiv e^2/4\pi$ is the electromagnetic fine-structure constant, and the factors in bracket represent the contributions from the $H^+$, $\delta_L^+$, $\delta_R^{++}$, $\delta_L^{++}$, and $W_R^{+}$ loops, respectively.\footnote{Had we kept the mixing angles $\theta_{1,2}$ there would be additional contributions and even the possibility to fine-tune $\Gamma_{\chi \to \gamma \gamma} \simeq 0$, which could make possible a much lower $v_R$ scale.}
We refrain from discussing other possible decay channels such as $\chi\to \nu\nu$ or $\chi\to e^+ e^-$ because they are less important in the mass range of interest.

\begin{figure}[tb]
    \centering
    \includegraphics[width=0.47\textwidth]{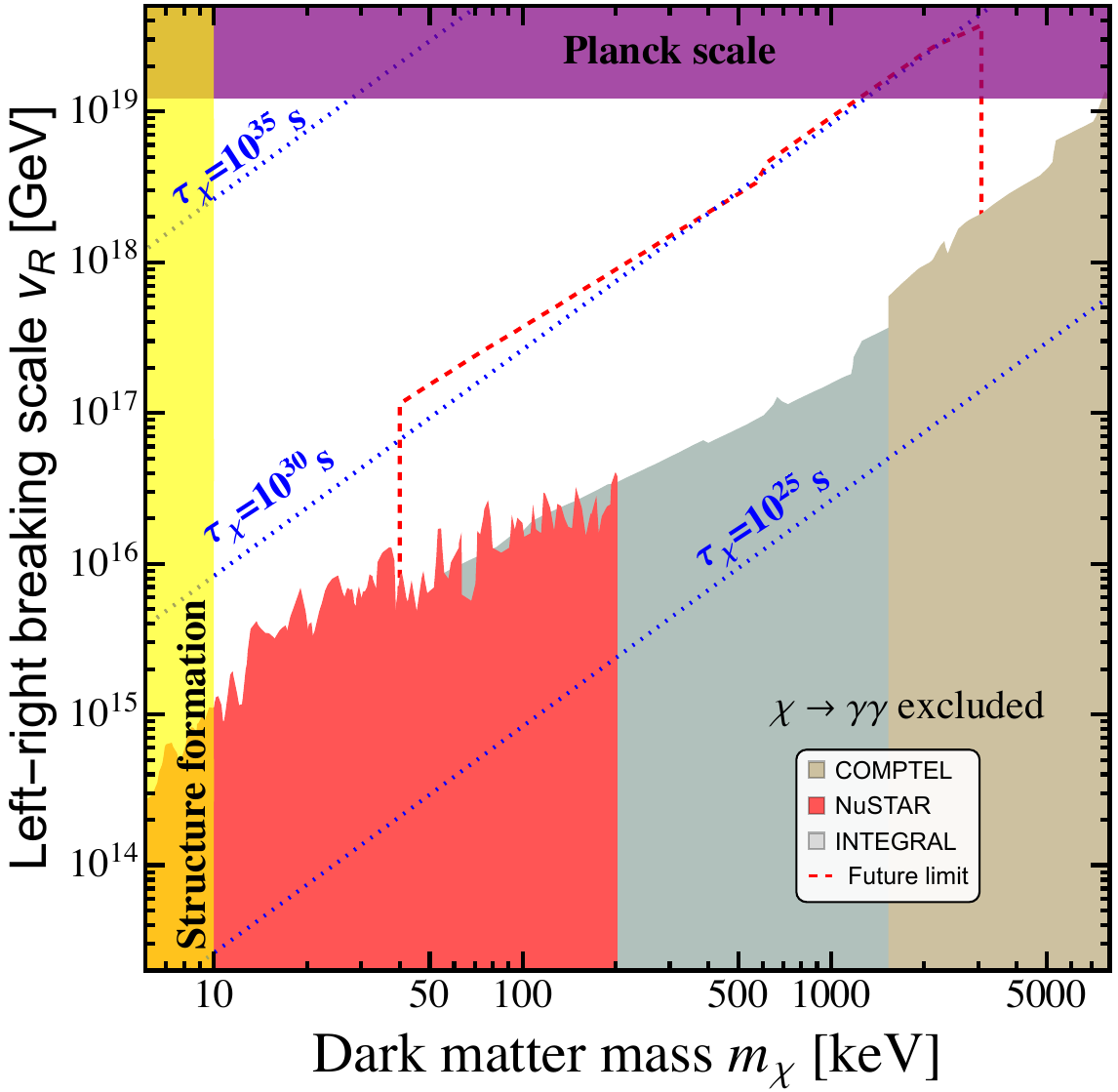}
    \caption{Allowed parameter space (unshaded) for the DM mass $m_\chi$ as a function of the LR breaking scale $v_R$. The  shaded regions are excluded.  X-ray constraints are obtained from  NuSTAR~\cite{Ng:2019gch,Roach:2022lgo} (red), INTEGRAL~\cite{Calore:2022pks,Laha:2020ivk} (gray), and COMPTEL~\cite{Essig:2013goa} (brown).
The red dashed line represents future X-ray reach~\cite{Cooley:2022ufh,ODonnell:2024aaw}.
The yellow region corresponds roughly to the excluded region due to Lyman-alpha data, assuming a (semi)thermal DM production mechanism~\cite{Heeck:2017xbu}.
The purple shaded region on top corresponds to $v_R>M_{\rm Pl}$, where quantum gravity effects might invalidate our perturbative calculation. Shown in dotted blue  are contours of the DM lifetime $\chi\to\gamma\gamma$. 
} 
    \label{fig:mainplot}
\end{figure}

If $\chi$ makes up all DM in our Universe, the di-photon decay has to be enormously suppressed to satisfy the stringent experimental limits~\cite{ODonnell:2024aaw}.\footnote{\change{This turns out to be a much stronger requirement than simply the cosmological stability of the DM.}} This can be achieved by an extreme hierarchy $m_\chi\ll v_R$. The lower limit $m_\chi \gtrsim \unit[10]{keV}$ from structure formation~\cite{Heeck:2017xbu} forces $v_R$  above $\unit[10^{15}]{GeV}$ to satisfy X-ray line search limits~\cite{Essig:2013goa, Ng:2019gch,Laha:2020ivk, Roach:2022lgo, Calore:2022pks}, illustrated in Fig.~\ref{fig:mainplot}.
While such a high LR breaking scale is not unfamiliar from grand unified theories, it exacerbates the fine-tuning needed to make $\chi$ light. \change{On the other hand, for $v_R<M_{\rm Pl}$, we must have $m_\chi\lesssim$ 10 MeV to satisfy the gamma-ray constraints.}

\begin{figure}
    \centering
    \includegraphics[width=0.35\linewidth]{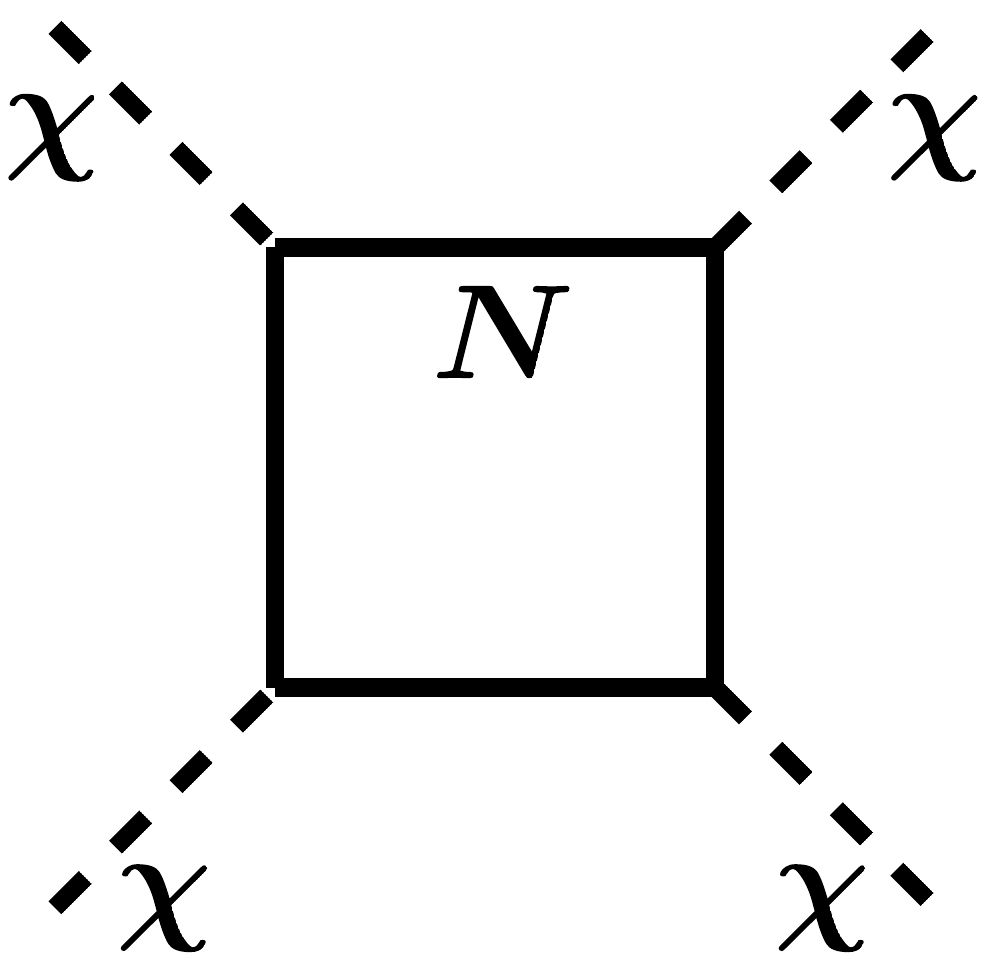}
    \caption{One-loop contribution to the quartic-coupling constant $\rho_1$ from the Yukawa coupling $f$ of Eq.~\eqref{eq:LagLR}. }
    \label{fig:r1RGE}
\end{figure}

We discuss the scalar potential and the conditions for stable parameters  $\rho_1,\theta_{1,2}\lll 1$ in appendices~\ref{sec:AppA} and~\ref{sec:App2}, respectively. $\rho_1$, one of the quartic couplings of $\Delta_R$, is particularly important in this setup:
\begin{align}
   \hspace{-1ex}  \rho_1\simeq \frac{m_\chi^2}{2v_R^2} \simeq 5\times 10^{-43}\left(\frac{m_\chi}{\unit[100]{keV}}\right)^2 \left(\frac{\unit[10^{17}]{GeV}}{v_R}\right)^2 \hspace{-0.5ex}.
\end{align}
It is chosen to be near zero at the scale $m_\chi$, and is thus highly sensitive to renormalization-group running effects. Loop corrections coming from bosonic loops will increase $\rho_1$, but the contributions from the RH neutrinos $N$ drive $\rho_1$ to \textit{negative} values, which would destabilize the vacuum. This behavior is of course completely analogous to the hierarchy problem and vacuum stability issue in the SM~\cite{Buttazzo:2013uya, Hiller:2024zjp}. If we assume for now that all non-SM bosons except $\chi$ have masses near the $v_R$ scale, $\rho_1$ remains constant up to the mass scale $m_N$ corresponding to the lightest $N$, and then picks up a negative contribution
\begin{align}
  \Delta \rho_1 \simeq - \frac{16 |f|^4}{(4\pi)^2}\log \left(\frac{v_R}{m_N}\right)
  \label{eq:Delta_rho1}
\end{align}
on the way to $v_R$, coming from the loop contribution to the quartic coupling shown in Fig.~\ref{fig:r1RGE}. If $\rho_1 + \Delta \rho_1 > 0$ up to $v_R$, it will remain positive even at larger scales due to the large positive contributions from the heavy bosons. But to ensure this, $N$ should either sit near, or ideally above, $v_R$ as well (to avoid $\rho_1$ running altogether), or have a tiny Yukawa coupling $f$, roughly $|f| < \sqrt{\pi} \rho_1^{1/4}$, which translates into $m_N \lesssim \sqrt{m_\chi v_R}$.
Here we assumed only one $N$; if instead we have three degenerate $N$ with same coupling $f$, the bound is marginally modified to $m_N \lesssim \sqrt{m_\chi v_R/\sqrt3}$, which makes little difference to our analysis.

Vacuum stability of this DM setup thus forbids the RH neutrino mass region
\begin{align}
     \sqrt{m_\chi v_R} \lesssim m_N \lesssim v_R \,.
\end{align}
Even lowering the masses of some of the charged scalars below $v_R$ to utilize their positive contributions to $\rho_1$ does not change this conclusion, or at least our numerical scans did not reveal any viable scenarios.
The new scalars can, however, cure the \textit{electroweak} vacuum stability problem, as shown in appendix~\ref{sec:App3}. Since this has little to no bearing on our DM phenomenology, we will not discuss it further.
Below, we discuss the two viable mass windows for $N$ and possible ways to generate the correct DM abundance, neutrino masses, and the baryon asymmetry of our Universe.

\section{Light right-handed neutrinos}\label{sec:lightN}

In the parameter region $m_N < \sqrt{m_\chi v_R} $, the RH neutrinos are far below the LR scale, but can still be far above the electroweak scale:
\begin{align}
    m_N \lesssim \unit[3\times 10^6]{GeV}\left( \frac{m_\chi}{\unit[100]{keV}}\right)^{\frac12}\left(  \frac{v_R}{\unit[10^{17}]{GeV}}\right)^{\frac12} .
    \label{eq:upper_bound_on_mN}
\end{align}
In this region of parameter space we can explain the active neutrino masses via type-I seesaw by adjusting the Dirac-mass matrix to be of order
\begin{align}
\begin{split}
    M_{\nu_D} &\simeq \sqrt{m_\nu m_N}\\
    &\simeq \unit[20]{MeV}\left( \frac{m_\nu}{\unit[0.1]{eV}}\right)^{\frac12}\left(  \frac{m_N}{\unit[3\times 10^{6}]{GeV}}\right)^{\frac12} .
\end{split}
\label{eq:seesaw_MD}
\end{align}
In addition, the type-II seesaw contribution could be sizeable despite the small $|f| < \sqrt{\pi} \rho_1^{1/4}$:
\begin{align}
m_\nu^{II} &\simeq \sqrt{2}  f v_L\\
&\simeq \unit[0.07]{eV}\left( \frac{m_\chi}{\unit[100]{keV}}\right)^{\frac12}\left(  \frac{\unit[10^{17}]{GeV}}{v_R}\right)^{\frac12}\left( \frac{v_L}{\unit{GeV}}\right) .\nonumber
\end{align}
Given the upper limit on $v_L$ of order GeV from electroweak precision measurements~\cite{Heeck:2022fvl,ParticleDataGroup:2024cfk}, there is a sizable region of parameter space in Fig.~\ref{fig:mainplot} where most or even all of $m_\nu$ comes from the type-II seesaw mechanism.  For very large $v_R$ and small $m_\chi$, type-I seesaw is going to be dominant though.

If the Universe reached temperatures around $v_R$, the scalar DM would have been thermalized through the RH gauge interactions. Assuming all triplet partners to have masses around $v_R$, the real scalar $\chi$ would have frozen out at a temperature $T_f$ just below $v_R$ while being relativistic, giving an abundance today~\cite{Kolb:1990vq}
    \begin{align}
        \Omega_\chi h^2 \simeq 0.12 \left(\frac{106.75}{g_*(T_f)}\right)\left(\frac{m_\chi}{\unit[170]{eV}}\right) .
        \label{eq:relativistic_freeze-out}
    \end{align}
For $m_\chi \sim \unit[170]{eV}$, this would match the observed DM abundance $\Omega h^2 \simeq 0.12$~\cite{Planck:2018vyg} but be \emph{hot} DM, completely excluded. For $m_\chi > \unit[170]{eV}$, it would overclose the Universe, also excluded. The same problem occurs for the LR keV sterile-neutrino DM case -- 
 only the abundance is increased by $2\times 3/4$ due to the fermionic degrees of freedom. Let us discuss a few possible solutions:
 \begin{enumerate}
 
     \item Entropy dilution through $g_*$: The DM abundance in Eq.~\eqref{eq:relativistic_freeze-out} can equal the observed one even for $m_\chi \gg \unit[170]{eV}$ if we assume additional degrees of freedom at the time of DM freeze-out, i.e.~an increased $g_*(T_f)$. This solution is not particularly appealing since it requires the addition of many more particles beyond those in our LR model~\cite{Bezrukov:2009th}.

     \item Entropy dilution from late decays: The relatively light RH neutrinos will be in thermal equilibrium around $v_R$ and freeze out relativistically around a similar temperature  $T_f$ as $\chi$ since both couple to the $SU(2)_R $ gauge bosons.
     If the lifetime of one of the $N$ is sufficiently long, it can dominate the Universe for a while once $T\lesssim m_N$, and then decay, thus reheating the Universe to a temperature~\cite{Scherrer:1984fd}
    \begin{align}
        \qquad  T_\text{reh} &\simeq 0.8\, (g_*(T_\text{reh}))^{-\tfrac{1}{4}} \sqrt{\Gamma_N M_{\rm Pl}}\\
        &\simeq \unit[700]{GeV} \left(\frac{g_*(T_\text{reh})}{106.75}\right)^{-\frac14}\left(\frac{\unit[10^{-12}]{s}}{\tau_N}\right)^{\frac12} ,
    \end{align}
    which effectively dilutes the DM temperature because $N$ does not decay into DM -- unlike in the keV sterile neutrino case~\cite{Nemevsek:2022anh,Nemevsek:2023yjl}. This can give cold DM with the right abundance as long as $m_N$ and $\tau_N$ are picked correctly so that enough entropy is produced (long $N$ lifetime) while keeping the reheating temperature above MeV for successful Big Bang nucleosynthesis (short $N$ lifetime).
    Concretely,  $\Omega_\chi$ is reduced by a factor~\cite{Bezrukov:2009th,Nemevsek:2012cd}
    \begin{align}
        \quad\ \  S \simeq 80 \left(\frac{106.75}{g_*(T_\text{reh})}\right)^{\tfrac34}\left(\frac{\tau_N}{\unit[10^{-12}]{s}}\right)^{\tfrac12} \left(\frac{m_N}{\unit[10^{7}]{GeV}}\right) ,
    \end{align}
    thus allowing for $m_\chi > \unit[170]{eV}$, 
    and its temperature is reduced by a factor $S^{1/3}$, making it much colder. As a result, the Lyman-$\alpha$ limits in this case are those of a thermal relic~\cite{Bezrukov:2009th}, excluding masses below $\unit[4]{keV}$~\cite{Boulebnane:2017fxw}. We then need at least $S\gtrsim 25$.
    The required $N$ lifetime is far longer than expected from the seesaw relation in Eq.~\eqref{eq:seesaw_MD}, so we need to assume that one of the three $N$ has much smaller Yukawa couplings, leading to a longer lifetime and one effectively massless active neutrino compatible with current data.\footnote{The type-II seesaw contribution can and will lift the lightest neutrino mass.}\\
    Despite the upper bound on $m_N$ from Eq.~\eqref{eq:upper_bound_on_mN}, there is even a region of parameter space in which all $N$ decay before sphaleron decoupling, allowing for leptogenesis. The small masses compared to the Davidson-Ibarra bound~\cite{Davidson:2002qv} imply generically too small a CP asymmetry, but two or more degenerate $N$ could lead to resonant leptogenesis to circumvent this problem~\cite{Pilaftsis:2003gt,Bezrukov:2012as}. Even thermal leptogenesis could work all the way down to $10^6$ GeV if we consider flavor effects~\cite{Moffat:2018wke}.

     \item Reheating temperature $T_\text{reh} $ below $v_R$: If we obtain too much DM for $T_\text{reh} > v_R$ and too little DM for $T_\text{reh} \ll v_R$ -- seeing as $\chi$ only interacts weakly with the SM through $v_R$-suppressed operators -- there must be a sweet spot $T_\text{reh}< v_R$ where the correct DM abundance is produced for any $m_\chi$. In this UV freeze-in~\cite{Elahi:2014fsa} case, DM is produced with momenta $\langle p \rangle\sim 3\,T$, so masses below $\sim\unit[10]{keV}$ are excluded by Lyman-$\alpha$~\cite{Boulebnane:2017fxw}.\\
     Since the RH neutrinos have by and large the same interactions as $\chi$, they should generically be produced by the same amount at $T_\text{reh}$. They stay relativistic until $T\sim m_N$ and then decay. For seesaw-inspired $M_{\nu_D}$ values, the decay will be too fast to lead to a period of matter domination. The lepton asymmetry produced in their decay (subsequently converted to a baryon asymmetry by electroweak sphalerons) is generically too small to explain the matter asymmetry since the small $m_N$ suppresses the CP asymmetry $\epsilon$. However, \emph{resonant} leptogenesis~\cite{Pilaftsis:2003gt}, i.e.~two or more almost degenerate RH neutrinos, could work in this scenario~\cite{Bezrukov:2012as}, at the price of additional fine-tuning.
 \end{enumerate}

Overall, the light $N$ window of the scalar-DM parameter space offers numerous ways to simultaneously explain neutrino masses, the DM abundance, and even leptogenesis. Let us now discuss the heavy-$N$ scenario.

\section{Heavy right-handed neutrinos}

With $m_N > v_R$ and $v_R \gtrsim \unit[10^{15}]{GeV}$ from Fig.~\ref{fig:decaygg}, the type-I seesaw contribution to active neutrino masses is suppressed:
\begin{align}
\begin{split}
    m_\nu^I %&\simeq M_{\nu_D}^2/m_N\\
    &\lesssim \unit[0.04]{eV}\left( \frac{M_{\nu_D}}{\unit[200]{GeV}}\right)^{2}\left(  \frac{\unit[ 10^{15}]{GeV}}{m_N}\right) ,
\end{split}
\end{align}
and could at most account for the two lightest neutrino masses unless we push the Yukawa couplings near the non-perturbative regime.
In this region of parameter space, neutrino masses have to come from type-II seesaw, and since $f$ is of order one in this section we require a tiny $v_L \sim m_\nu$.

For $m_N > v_R$, we can still obtain the correct DM abundance through UV freeze-in or by severely increasing $g_*(T_f)$, but the late-entropy solution is no longer viable because the $N$ will freeze-out non-relativistically and thus cannot dump enough entropy into the SM bath to solve the overproduction problem. The minimal LR model does not contain any other particles that could lead to late entropy production.
In that case, the heavy $N$ were never produced and thus cannot be used for leptogenesis. 
Upon introduction of additional particles, one could arrange to produce $N$ despite the low temperature, e.g.~via inflaton decays~\cite{Barman:2021tgt}   or bubble collisions~\cite{Cataldi:2024pgt}, which could then lead to leptogenesis. In this case it is important to note that the large coupling $f$ of $\chi$ to $N$ means that every $N$ decay/production comes with an $\mathcal{O}(1)$ number of $\chi$ particles, e.g.
\begin{align}
    \frac{\Gamma (N\to L H \chi)}{\Gamma (N\to LH)} \simeq \frac{|f|^2}{2\pi^2} \log \left[\frac{M_N}{m_\chi}\right],
\end{align}
enhanced by an infrared divergence since the virtual $N$ goes on shell for $m_\chi \to 0$ (and $p_\chi\to 0$, so the $\chi$ will be produced with very small momentum~\cite{Heeck:2017xbu}).
We will omit a discussion of such scenarios because they go beyond the minimal LR setup.

\section{Conclusion}

The left-right model has been a popular extension of the SM for half a century. While it naturally explains neutrino masses and can accommodate the baryon asymmetry of the Universe, dark matter has proven elusive for a long time. A decade or so ago, it was noticed~\cite{Nemevsek:2012cd} that one of the right-handed neutrinos within the LR model can be tuned light enough to be stable on cosmological timescales, allowing for it to be viable DM.
Here, we showed that the LR model contains yet another viable DM candidate in the form of the LR-breaking Higgs. With admittedly severe fine-tuning, it can be made light and stable enough to give DM. Vacuum stability imposes non-trivial constraints on the RH neutrino masses in this scenario, which however still allow for successful neutrino masses and leptogenesis. This DM candidate likely remains viable when we embed the LR model into Pati-Salam or even $SO(10)$, providing a completely different way to get DM there, as opposed to the WIMP-like stable DM candidates proposed before~\cite{Heeck:2015qra,Garcia-Cely:2015quu,Ferrari:2018rey}. \change{Observing an X-ray or gamma-ray line signal from DM-rich regions would be a possible indication of this DM candidate. Finding a $W_R$ boson at the LHC or future collider would exclude this DM model.}

\section*{Acknowledgment}

We thank Yongchao Zhang for useful discussion. B.D. thanks the HEP group at UVA for a seminar visit, where this work was initiated. B.D.~and A.T. wish to thank the
Center for Theoretical Underground Physics and Related
Areas (CETUP*) and the Institute for Underground Science at SURF for their hospitality and for providing a
stimulating environment, where part of this work was done. The work of B.D. was partly
supported by the U.S. Department of Energy under grant
No. DE-SC0017987. 
The work of J.H.~was supported by the U.S. Department of Energy under Grant No. DE-SC0007974.

%\begin{widetext}
\onecolumngrid
\appendix
\section{Scalar Sector}
\label{sec:AppA}
We analyze the Higgs potential of the model to see if a light $\Re(\Delta_R^0)$ is viable. The most general renormalizable Higgs potential involving $\Phi$, $\Delta_L$ and $\Delta_R$ fields is given by~\cite{Deshpande:1990ip} 
\begin{align}
V (\Phi,\Delta_{L,R}) = & -\mu_1^2 {\rm Tr}(\Phi^{\dagger} \Phi)-\mu_2^2\left[{\rm Tr}(\Phi^{\dagger} \tilde{\Phi}+{\rm Tr}(\Phi \tilde{\Phi}^{\dagger})\right]-\mu_3^2 \left[ {\rm Tr}(\Delta_L^{\dagger} \Delta_L) + {\rm Tr}(\Delta_R^{\dagger} \Delta_R) \right]+\lambda_1 {\rm Tr}(\Phi^{\dagger} \Phi)^2 \notag \\
& +\lambda_2\left[{\rm Tr}(\Phi^{\dagger} \tilde{\Phi})^2+{\rm Tr}(\Phi \tilde{\Phi}^{\dagger})^2\right]  + \lambda_3 {\rm Tr}(\Phi^{\dagger} \tilde{\Phi}) {\rm Tr}(\Phi \tilde{\Phi}^{\dagger}) +\lambda_4 {\rm Tr}(\Phi^{\dagger} \Phi)\left[{\rm Tr}(\Phi \tilde{\Phi}^{\dagger})+{\rm Tr}(\Phi^{\dagger} \tilde{\Phi})\right]  \notag\\
& +\rho_1 \left[ {\rm Tr}(\Delta_L^{\dagger} \Delta_L)^2 + {\rm Tr}(\Delta_R^{\dagger} \Delta_R)^2 \right] +\rho_2 \left[ {\rm Tr}(\Delta_L \Delta_L) {\rm Tr}(\Delta_L^{\dagger} \Delta_L^{\dagger}) +  {\rm Tr}(\Delta_R \Delta_R) {\rm Tr}(\Delta_R^{\dagger} \Delta_R^{\dagger}) \right]   \notag \\ 
& + \rho_3 {\rm Tr} (\Delta_L \Delta_L^{\dagger}) {\rm Tr}(\Delta_R \Delta_R^{\dagger}) + \rho_4 \left[{\rm Tr} (\Delta_L \Delta_L) {\rm Tr} (\Delta_R^{\dagger} \Delta_R^{\dagger}) +  {\rm Tr} (\Delta_L^\dagger \Delta_L^\dagger) {\rm Tr} (\Delta_R \Delta_R) \right]  \notag \\
&+ \alpha_1 {\rm Tr}(\Phi^{\dagger} \Phi) \left[ {\rm Tr}(\Delta_L^{\dagger} \Delta_L) + {\rm Tr}(\Delta_R^{\dagger} \Delta_R) \right] +\left[\alpha_2 \left\{ {\rm Tr} (\tilde{\Phi} \Phi^\dagger) {\rm Tr}(\Delta_L^{\dagger} \Delta_L) +  {\rm Tr} (\tilde{\Phi}^{\dagger} \Phi) {\rm Tr}(\Delta_R^{\dagger} \Delta_R) \right\} +\text { H.c. }\right] \notag \\
&+\alpha_3 \left[ {\rm Tr}(\Phi \Phi^{\dagger} \Delta_L \Delta_L^{\dagger}) + {\rm Tr} (\Phi^{\dagger} \Phi \Delta_R \Delta_R^{\dagger}) \right]  + \beta_1 \left[ {\rm Tr} (\Phi \Delta_R \Phi^\dagger \Delta_L^\dagger) + {\rm Tr} (\Phi^\dagger \Delta_L \Phi \Delta_R^\dagger) \right]  \notag \\ 
&+ \beta_2 \left[ {\rm Tr} (\tilde{\Phi} \Delta_R \Phi^\dagger \Delta_L^\dagger) + {\rm Tr} (\tilde{\Phi}^\dagger \Delta_L \Phi \Delta_R^\dagger) \right] + \beta_3 \left[ {\rm Tr} (\Phi \Delta_R \tilde{\Phi}^\dagger \Delta_L^\dagger) + {\rm Tr} (\Phi^\dagger \Delta_L \tilde{\Phi} \Delta_R^\dagger) \right]\, .
\label{eq:pot}
\end{align}
Here all the couplings except for $\alpha_2$ are made real by field redefinitions~\cite{Deshpande:1990ip}. 
Furthermore, for simplicity, we take the phase of $\alpha_2$ to be zero such that there is no CP-violation in the scalar sector and no mixing between scalars and pseudoscalars. Moreover, we take all the VEVs to be real. After inserting the VEVs, we obtain the following conditions for the potential to be an extremum:
\begin{align}
  -\mu_1^2 + k_{+}^2 \lambda_1+2 \kappa_1 \kappa_2 \lambda_4 + \frac{1}{2 k_{-}^2}\left[2 v_L v_R\left(\beta_2 \kappa_1^2-\beta_3 \kappa_2^{2}\right)+\left(v_L^2+v_R^2\right)\left(\alpha_1 k_{-}^2-\alpha_3 \kappa_2^{2}\right)\right] =&\ 0 ,\\
 - \mu_2^2 + \kappa_1 \kappa_2 \left(2 \lambda_2+\lambda_3\right)+ \frac{1}{2}\lambda_4 k_{+}^2 + \frac{1}{4 k_{-}^2}{\left[v_L v_R\left(\beta_1 k_{-}^2-2 \kappa_1 \kappa_2 \left(\beta_2-\beta_3\right)\right)+\left(v_L^2+v_R^2\right)\left(2 \alpha_2 k_{-}^2+\alpha_3 \kappa_1 \kappa_2\right)\right]}  =&\ 0 ,\\
-\mu_3^2 + \frac{1}{2} \left[\alpha_1 k_{+}^2+4 \alpha_2 \kappa_1 \kappa_2 + \alpha_3 \kappa_2^{2} + 2 \rho_1 v_R^2+2 \rho_1\left(v_L^2+v_R^2\right)\right] =&\ 0 ,\\
\beta_2 \kappa_1^2 + \beta_1 \kappa_1 \kappa_2 +\beta_3 \kappa_2^{2} - v_L v_R\left(2 \rho_1-\rho_3\right) =&\ 0  . \label{eq:seesawHiggs}
\end{align}
The last equation, Eq.~\eqref{eq:seesawHiggs}, is the VEV seesaw relation.  Here and in what follows we take $\kappa_1 \geq \kappa_2$ without loss of generality and define
\begin{equation}
    \kappa_{\pm}^2 \equiv \kappa_1^2 \pm \kappa_2^2\,.
\end{equation}
Using the minimization conditions to eliminate $\mu_i^2$, we construct the mass matrix for the neutral Higgs bosons in the basis $\{ \phi_1^{0r}, \phi_2^{0r}, \delta_R^{0r}, \delta_L^{0r}  \}$ by expanding the potential about the minimum in VEVs to quadratic order. Here the basis $\{ \phi_1^{0}, \phi_2^{0}, \delta_R^{0} , \delta_L^{0} \}$ consisting of complex fields is broken up into real and imaginary components; for instance $\phi_1^0 = (\phi_1^{0r} + i \phi_1^{0i})/\sqrt{2}$. 
To identify the SM-like Higgs, we use the following orthogonal matrix to transform the original basis to an intermediate flavor-diagonal basis:
\begin{equation}
    O_r M_r^2 O_r^T = \tilde{M}_r^2 \, ,
\end{equation}
where $M_r^2$ is the mass matrix in the $\{ \phi_1^{0r}, \phi_2^{0r}, \delta_R^{0r}, \delta_L^{0r}  \}$ basis and $O_r$ is given by
\begin{equation}
  O_r =  \begin{pmatrix}
        \kappa_1/\kappa_+ & \kappa_2/\kappa_+ & 0 & 0\\
       - \kappa_2/\kappa_+ & \kappa_1/\kappa_+ & 0 & 0 \\
        0 & 0 & 1 & 0 \\
        0 & 0 & 0 & 1
    \end{pmatrix} .
    \label{eq:rotatedbasis}
\end{equation}
The symmetric mass matrix in the rotated basis $\{ h_1^{0r}, h_{2}^{0r}, \delta_R^{0r}, \delta_L^{0r}\}$ with $v_L\simeq0$ (for simplicity) then reads 
\begin{equation}
  \tilde{M}_r^2 =
    \begin{pmatrix}
      2 \lambda_1 \kappa_+^2 + 8 \lambda_4 \kappa_1 \kappa_2 + 8 (2\lambda_2 + \lambda_3) \frac{\kappa_1^2 \kappa_2^2}{\kappa_+^2}  &  \bullet & \bullet & \bullet \\[5pt]
      %%%
      2 \lambda_4 \kappa_-^2 + 4 (2\lambda_2 + \lambda_3) \kappa_1 \kappa_2 \frac{\kappa_-^2}{\kappa_+^2} & 2 (2\lambda_2 + \lambda_3) \frac{\kappa_-^4}{\kappa_+^2} + \alpha_3 \frac{\kappa_+^2}{2 \kappa_-^2} v_R^2 & \bullet  & \bullet \\[5pt]
      %%%
      \left\{ \alpha_1 \kappa_+^2 + (4 \alpha_2 \kappa_1 + \alpha_3 \kappa_2) \kappa_2 \right\} \frac{v_R}{\kappa_+} & (2 \alpha_2 \kappa_-^2 + \alpha_3 \kappa_1 \kappa_2) \frac{v_R}{\kappa_+} & 2 \rho_1 v_R^2 & \bullet \\[5pt]
      %%%
      0 & \frac{1}{2\kappa_1} (\beta_1 \kappa_1 + 2 \beta_3 \kappa_2) \kappa_+ v_R & 0 & \frac{1}{2} (\rho_3 - 2\rho_1) v_R^2
    \end{pmatrix} .
    \label{eq:Neutmat}
\end{equation}
The $\delta_L^{0r}$ field decouples in the limit $\beta_i \simeq 0$ or $\beta_1 = 2\beta_3 \kappa_2/\kappa_1$. Note that such a choice is fully consistent with $v_L \simeq 0$ as evident by the seesaw relation of Eq.~\eqref{eq:seesawHiggs}. The field $h_1^{0r}$ is the SM-like Higgs boson in the limit of small mixing with other states. 
The remaining $3\times3$ matrix can be further diagonalized to obtain the mass eigenvalues of the real scalar fields. Here we wish to identify an additional state, i.e., $\delta_R^{0r}$ light such that it is a DM candidate.  We use the following transformation matrix to relate the flavor-diagonal basis $\{ h_1^{0r}, h_{2}^{0r}, \delta_R^{0r} \}$ and mass eigenstates of the $3\times3$ real scalar matrix. 
\begin{equation}
    \begin{pmatrix}
    h\\
    H \\
    \chi
\end{pmatrix} = 
\begin{pmatrix}
    \cos\theta_1 & 0 & \sin\theta_1 \\
    0 & 1 & 0 \\
    -\sin\theta_1 & 0 & \cos\theta_1 &
\end{pmatrix} 
\begin{pmatrix}
   1  & 0 & 0 \\
    0 & \cos\theta_2 & \sin\theta_2 \\
   0 & -\sin\theta_2 & \cos\theta_2 &
\end{pmatrix} 
\begin{pmatrix}
    h_1^{0r} \\
    h_{2}^{0r} \\
    \delta_R^{0r}
\end{pmatrix} .
\label{eq:mixingmatix}
\end{equation}
Here, $\sin\theta_{1,2}$ denotes the mixing between $\chi$ with $h$ and $H$. We ignored the mixing between $h$ and $H$ since they are of order ${\cal O}(1/v_R^2)$, thus heavily suppressed. In the limit of small mixing angles, the DM mass is given by Eq.~\eqref{eq:DM}, 
where the mixing angles are given by 
\begin{align}
    \theta_1 = \frac{\ C_1 \frac{v_R}{\kappa_+}}{m_h^2-m_{\chi}^2} \, ,\hspace{8mm}
    %%%
    \theta_2 = \frac{\left( 2 \alpha_2 \kappa_-^2 + \alpha_3 \kappa_1 \kappa_2\right) \frac{v_R}{\kappa_+}  }{m_H^2 - m_\chi^2} \, , \label{eq:sint2}
\end{align}
with $C_1 = \alpha_1 \kappa_+^2 + (4 \alpha_2 \kappa_1 + \alpha_3 \kappa_2) \kappa_2$.
In order for the mass of $\chi$ to be at or below the electroweak scale and its mixing with other Higgs fields negligible, one needs
$\rho_1 \sim {\cal O}(1/v_R^2)$ and 
\begin{equation}
    \alpha_1 \simeq \frac{\alpha_3 \kappa_2^2}{\kappa_-^2} \, , \hspace{10mm} \alpha_2 \simeq - \frac{\alpha_3 \kappa_1 \kappa_2}{2 \kappa_-^2} \, .
    \label{eq:a1a2}
\end{equation}

In the pseudoscalar sector, we can make a similar rotation from the original basis $\{ \phi_1^{0i}, \phi_2^{0i}, \delta_R^{0i}, \delta_L^{0i} \}$ to the mass basis $\{ A, G_L^{0}, G_R^{0},\delta_L^{0i}  \}$, where $G_{L}^{0}$ and $G_R^0 (\equiv \delta_R^{0i})$ are the Goldstone modes associated with neutral gauge bosons $Z_{L,R}$ with $G_L^0 \equiv (-\kappa_1 \phi_1^{0i} + \kappa_2 \phi_2^{0i})/\kappa_+$.
The mass of the physical pseudoscalars (in the limit $\beta_i \simeq 0$) then read as
\begin{equation}
    m_{A}^2 = - 2 (2 \lambda_2 - \lambda_3) \kappa_+^2 + \frac{1}{2} \alpha_3 \frac{\kappa_+^2}{\kappa_-^2} v_R^2 \, , \hspace{5mm} m_{\delta_L^{0i}}^2 = \frac{1}{2} (\rho_3 - 2\rho_1) v_R^2 \,.
\end{equation}
The singly charged scalars $\{ \phi_1^+, \phi_2^+, \delta_R^+, \delta_L^+ \}$ can be also rotated to the mass basis $\{ H^+, G_L^{+}, G_R^{+}, \delta_L^+ \}$. Here $G_{L,R}^{+}$ are the Goldstone modes associated with neutral gauge bosons $W_{L,R}^{+}$ and given by
\begin{align}
    G_L^+ = -\frac{-\kappa_1 \phi_1^+ + \kappa_2 \phi_2^+}{\kappa_+} \, , \hspace{10mm} G_R^+ = \frac{-\kappa_-^2 \phi_1^+ + \sqrt{2} \kappa_2 v_R \delta_R^+}{\sqrt{\kappa_-^4 + 2 \kappa_2^2 v_R^2}} \,. 
\end{align}
The orthogonal state to $\{ G_L^+ , G_R^+\}$ corresponds to the physical field $H^+$. The masses of $H^+$ and $\delta_L^+$ fields are 
\begin{equation}
    m_{H^+}^2 = \frac{1}{4} \alpha_3 \left\{ \kappa_-^2 + 2 \frac{\kappa_+^2}{\kappa_-^2} v_R^2 \right\} \, , \hspace{5mm} m_{\delta_L^+}^2 = \frac{1}{2} \left[ \alpha_3 \kappa_-^2 -  v_R^2 (2\rho_1-\rho_3) \right] \,  . 
\end{equation}
The masses of the doubly-charged scalars $(\delta_R^{++},\delta_L^{++})$ in the limit $\beta_i\simeq0$  read as
\begin{equation}
    m_{\delta_R^{++}}^2 = \frac{1}{2} \alpha_3 \kappa_-^2 + 2 \rho_2 v_R^2 \, , \hspace{5mm}m_{\delta_L^{++}}^2 = \frac{1}{2} \left[ \alpha_3 \kappa_-^2 - v_R^2 (2\rho_1-\rho_3) \right]  .
    \label{eq:massdouble}
\end{equation}
For completeness we also show the relevant couplings of $\delta_R^0 \equiv \chi$ in the mass basis in Tab.~\ref{tab:couplings} in the limit of $\beta_i = v_L = 0$ and $v_R \gg \kappa_1, \kappa_2$. 
\begin{table}[h]
    \centering
    \begin{tabular}{|c|c|}
    \hline
      {\bf Interaction}   & {\bf couplings }  \\
      \hline
       $\chi^4$    &  $ \rho _1/4$ \\
       $\chi^3$ & $  v_R \rho _1- \theta_1 C_1 / (2\kappa_+)$  \\
       \hline
       %%%
       $\chi \chi h h$ &  $ C_1/(4 \kappa_+^2)$ \\
       $\chi \chi h$ & $  C_1/(2\kappa_+) +  \theta_1 v_R \left(3 \rho _1 -\alpha _1-\kappa _2 \left(4 \alpha _2 \kappa _1+\alpha _3 \kappa _2\right)/\kappa_+^2\right)$ \\
       %%%
       $\chi h h$ & $ v_R C_1 / (2 \kappa_+^2)$ \\
       %%%
       $\chi \chi H H$  & $C_2/(4 \kappa_+^2)$  \\
       %%%
       $\chi \chi H$ & $- \left(\alpha _3 \kappa _1 \kappa _2+2 \alpha _2 \kappa_-^2\right) \left( \kappa_+^2-2  \theta_1 v_R \kappa_+\right)/(2 \kappa_+^3)$ \\
       $\chi H H$ & $v_R C_2/(2 \kappa_+^2)$ \\
       %%%
       $\chi \chi \delta_{L}^{0r} \delta_{L}^{0r}$ & $  \rho _3/4$ \\
       $\chi \delta_{L}^{0r} \delta_{L}^{0r}$ & $  v_R \rho _3/2- \theta_1 C_1/(2\kappa_+)$ \\
       %%%
      $\chi \chi A A$ & $ C_2/(4 \kappa_+^2)$ \\
      %%%
      $\chi A A$ & $ v_R C_2/(2 \kappa_+^2)$ \\
      %%%
      $\chi \chi \delta_L^{0i} \delta_L^{0i}$ & $ \rho _3/4$ \\
      %%%
      $\chi \delta_L^{0i} \delta_L^{0i}$ & $ v_R \rho _3/2- \theta_1 C_1/(2\kappa_+) $ \\
      \hline
      %%%
      $\chi \chi H^+ H^-$ & $ \left(C_2 v_R^2+\kappa_-^4 \rho _1\right)/(2 v_R^2 \kappa_+^2)$ \\
     $ \chi H^+ H^-$ & $\left(2 C_2 v_R^2+\alpha _3 \left(\kappa _1^4-\kappa _2^4\right)+2 \kappa_-^4 \rho _1\right)/(2 v_R \kappa_+^2)$ \\
     %%%
     $\chi \chi \delta_L^+ \delta_L^-$ & $\rho _3/2$ \\
     %%%
     $\chi \delta_L^+ \delta_L^-$ & $- \theta_1 \kappa_+ \left(2 \alpha _1+\alpha _3\right)/2+ v_R \rho _3- 4  \theta_1 \alpha _2 \kappa _1 \kappa _2/\kappa_+$ \\
     \hline
     %%%
     $\chi \chi \delta_R^{++} \delta_R^{--}$ & $ \rho _1+2 \rho _2$ \\
     %%%
     $\chi \delta_R^{++} \delta_R^{--}$ & $2  v_R \left(\rho _1+2 \rho _2\right)- \theta_1 \left(\left(\alpha _1+\alpha _3\right) \kappa _1^2+4 \alpha _2 \kappa _2 \kappa _1+\alpha _1 \kappa _2^2\right)/\kappa_+$ \\
     %%%
     $\chi \chi \delta_L^{++} \delta_L^{--}$ & $ \rho _3/2$ \\
     $\chi \delta_L^{++} \delta_L^{--}$ & $ v_R \rho _3- \theta_1 \left(\left(\alpha _1+\alpha _3\right) \kappa _1^2+4 \alpha _2 \kappa _2 \kappa _1+\alpha _1 \kappa _2^2\right)/\kappa_+$ . \\
     \hline
    \end{tabular}
    \caption{Couplings of scalar DM $\chi$ in the limit of $\beta_i = v_L = 0$ and $v_R \gg \kappa_1, \kappa_2$ in the minimal LR model. Only the terms up to order $\mathcal{O}(\theta_1)$ and $\mathcal{O}(\theta_2)$ are kept. 
    Here, we define $C_2 = \left(\alpha _1+\alpha _3\right) \kappa _1^2-4 \alpha _2 \kappa _2 \kappa _1+\alpha _1 \kappa _2^2$. $C_1$ is defined through $\theta_1$ below Eq.~\eqref{eq:sint2} and vanishes upon substitution of Eq.~\eqref{eq:a1a2}.
    } 
    \label{tab:couplings}
\end{table}

\section{\change{Decay width of scalar DM to photons}}
\label{app:DMdecay}

\change{The decay of scalar dark matter $\chi$ to pair of photons is induced at one loop order through the gauge bosons $W_R^\pm$ and the charged scalars $\delta^{\pm\pm}_{L,R}$, $\delta_L^\pm$ and $H^\pm$ as shown in Fig.~\ref{fig:decaygg}. The decay width reads as \cite{Dev:2017dui}
\begin{equation}
\begin{aligned}
\Gamma\left(\chi \rightarrow \gamma \gamma\right)\simeq\frac{\alpha^2 m_{\chi}^3}{1028 \pi^3}  \frac{4}{v_R^2} \left \lvert\,  A_0(\tau_{H^{ \pm}})+ A_0(\tau_{\delta_L^{ \pm}}) +4  A_0(\tau_{\delta_R^{ \pm \pm}}) + 4  A_0(\tau_{\delta_L^{ \pm \pm}})  %+ \frac{\sqrt{2}}{v_{\mathrm{EW}}} \sum_{f=q, \ell} f_f N_C^f Q_f A_{1 / 2}\left(\tau_f\right)
+ A_1\left(\tau_{W_R^\pm}\right) \right|^2 \, ,
\end{aligned}
\end{equation}
and the loop functions are given by
\begin{align}
A_0(\tau) & \equiv-[\tau-f(\tau)] \tau^{-2} \,,\\
%A_{1 / 2}(\tau) & \equiv 2[\tau+(\tau-1) f(\tau)] \tau^{-2} \\
A_1(\tau) & \equiv-\left[2 \tau^2+3 \tau+3(2 \tau-1) f(\tau)\right] \tau^{-2}\,,
\end{align}
with $\tau_i=m_{\chi}^2 / 4 m_i^2$ and 
\begin{align}
f(\tau) \equiv\left\{\begin{array}{cc}
\arcsin ^2 \sqrt{\tau} & (\text { for } \tau \leq 1) \,,\\
-\frac{1}{4}\left[\log \left(\frac{1+\sqrt{1-1 / \tau}}{1-\sqrt{1-1 / \tau}}\right)-i \pi\right]^2 & (\text { for } \tau>1)\,.
\end{array}\right.
\end{align}
For heavy particles running inside the loop ($m_i\gg m_\chi$), these loop functions can be approximated to $A_0 (0) \approx 1/3$ and $A_1 (0) \approx -7$.}

\section{Theoretical Constraints}\label{sec:App2}
The two-body scalar scattering through contact interaction at tree level gives the following unitarity constraints on the quartic couplings~\cite{Chakrabortty:2016wkl}:
\begin{align}
& \lambda_1< 4 \pi / 3, ~~~~~ \left(\lambda_1+4 \lambda_2+2 \lambda_3\right)<4 \pi, ~~~~~ \left(\lambda_1-4 \lambda_2+2 \lambda_3\right)<4 \pi, ~~~~~ \lambda_4<4 \pi / 3, ~~~~~ \alpha_1<8 \pi, ~~~~~ \alpha_2<4 \pi,  \notag \\
&  \left(\alpha_1+\alpha_3\right)<8 \pi, ~~~~~ \rho_1<4 \pi / 3, ~~~~~ \left(\rho_1+\rho_2\right)<2 \pi, ~~~~~ \rho_2<2 \sqrt{2} \pi , ~~~~~ \rho_3 < 8 \pi , ~~~~~ \rho_4 < 2 \sqrt{2} \pi.
\label{eq:unit}
\end{align}
The quartic terms $V^{(4)} (\Phi, \Delta_R)$ form a vector space spanned by the real-valued vectors $x^T = \{\kappa_1^2, \kappa_2^2, v_R^2 \}$, which can be written as $V^{(4)} = x^T \hat{\lambda}\ x$ with
\begin{equation}
  \hat{\lambda} = 
  \frac{1}{4} \begin{pmatrix}
    \lambda_1  &~~~  \lambda' &~~~ \alpha_1/2 \\
   \lambda' &~~~ \lambda_1 &~~~ (\alpha_1 + \alpha_3)/2 \\
   \alpha_1/2 &~~~ (\alpha_1 + \alpha_3)/2 &~~~ \rho_1
    \end{pmatrix} .
    \label{eq:bound}
\end{equation}
Here, $\lambda' = \lambda_{1} + 4 \lambda_{2} + 2 \lambda_{3}$. The necessary conditions for the boundedness can be derived from the co-positivity of the real symmetric matrix $\hat{\lambda}$ of Eq.~\eqref{eq:bound}, which are~\cite{HADELER198379, Klimenko:1984qx, Chakrabortty:2013mha}:
\begin{align}
    &\lambda_1 \geq 0, ~~~~~ \rho_1 \geq 0, ~~~~~  2\lambda_2 + \lambda_3 \geq - \lambda_1, ~~~~~ \alpha_1 \geq  - 2 \sqrt{\lambda_1 \rho_1}, ~~~~~ \alpha_1 + \alpha_3 \geq  - 2\sqrt{\lambda_1 \rho_1} \, , \notag \\
 &  \lambda' \sqrt{\rho_1} + \frac{1}{2}(\alpha_1+ \alpha_3)\sqrt{\lambda_1} + \frac{1}{2}\alpha_1 \sqrt{\lambda_1} + \sqrt{\lambda_1^2\, \rho_1} \geq 0 ~~~~  \text{or}~~~ \text{det} \hat{\lambda} \geq 0 \, .
    \label{eq:bound2}
\end{align}
Note that the inclusion of $v_L$ leads to a symmetric matrix of order four. To determine co-positivity of such a matrix one would need to adjudge eight different cases depending on the sign distributions of the off-diagonal elements~\cite{Chakrabortty:2013mha}.
The full set of analytical expressions for vacuum stability that yield the desirable vacuum expectation values ensuring correct symmetry breaking to SM is analyzed in Ref.~\cite{Dev:2018xya, Chauhan:2019fji}.

\section{Renormalization Group Equations}\label{sec:App3}
Here we discuss the difficulty in maintaining a stable light-$\chi$ solution due to renormalization group equations (RGEs). For the potential to remain bounded from below, the quartic coupling must  satisfy the conditions outlined in Eq.~\eqref{eq:bound2}. Specifically, the quartic coupling $\rho_1 \chi^4$ needs to be exceedingly small, making it highly sensitive to any new interaction. Lowering the masses of one or more RH neutrinos below $v_R$ can potentially destabilize the vacuum, as the Yukawa coupling $f$ could drive the quartic coupling $\rho_1$ to negative values. The beta function for $\rho_1$ is given by $\beta(\rho_1) \propto - 16 f^4$ and picks up the negative contribution as illustrated in Eq.~\eqref{eq:Delta_rho1}.
A possible mitigation strategy is to keep scalar particles at or below $m_N$ to offset the negative $\Delta\rho_1$ from $N$ with a positive one from a scalar. However, we did not find any consistent solution up to two-loop that satisfies both the boundedness as well the vacuum stability conditions. The two-loop beta functions were generated with the {\tt PyR@TE} package~\cite{Sartore:2020gou} and were numerically solved in {\tt Mathematica} by introducing various scalar particles of the LR theory below $m_N$.

Alternatively, by choosing a tiny Yukawa coupling $f$, roughly $|f|< \sqrt{\pi} \rho_1^{1/4}$ or equivalently $m_N \lesssim \sqrt{m_\chi v_R}$, we can ensure that $\rho_1$ remains positive up to the $v_R$ scale. For instance, a benchmark point with $m_\chi = 1$ MeV, $v_R = \unit[10^{17}]{GeV}$, and $f \lesssim 6 \times 10^{-11}$ yields a consistent solution. For this benchmark, the corresponding mass of RH neutrino is $m_N \lesssim 8.5 \times 10^{6}$ GeV and the value $v_L \sim 0.59$ GeV ensures a neutrino mass of $\unit[0.1]{eV}$ via the type-II seesaw mechanism. The phenomenology of such a parameter region ($f\ll1$) is discussed in detail in Sec.~\ref{sec:lightN}.

Above the scale $v_R$, the gauge couplings of $SU(2)_R$ and $U(1)_{B-L}$ as well as all other scalar fields can contribute to the RGE running of $\rho_1$, with the one-loop beta function given by
\begin{align}
   \beta^{(1)}(\rho_{1}) =&
+ 28 \rho_{1}^{2}
+ 16 \rho_{1} \rho_{2}
+ 16 \rho_{2}^{2}
+ 3 \rho_3^{2}
+ 4 \alpha_{1}^{2}
+ 4 \alpha_{1} \alpha_{3}
+ 16 \left|{\alpha_{2}}\right|^{2}
+ 2 \alpha_{3}^{2} \notag \\
&- 12 g_{BL}^{2} \rho_{1}
- 24 g_R^{2} \rho_{1}
+ 6 g_{BL}^{4}
+ 12 g_{BL}^{2} g_R^{2}
+ 9 g_R^{4}
+ 8 \rho_{1} {\rm Tr}\left(f^\dagger f \right)
- 16 {\rm Tr}\left(f^\dagger f f^\dagger f \right) \, . \label{eq:rgeR1}
\end{align}
The gauge couplings $(g_R,g_{BL})$  originate from the $SU(2)_R$ and $U(1)_{B-L}$ gauge bosons, while the quartic couplings $\rho_{1,2,3}$ ($\alpha_{1,2,3}$) arise from the fields $\Delta_{L,R}$ ($\phi$ and $\Delta_{L,R}$) at the LR breaking scale. Notably, the negative contribution from the Yukawa coupling $f$, arising from fermion field $N$, can be made to cancel by the bosonic contribution, allowing $\rho_1$ to remain small. Consequently, the DM mass can be kept at or below MeV scale across all scales. 

\begin{figure}[t!]
    \centering
    \includegraphics[width=0.45\linewidth]{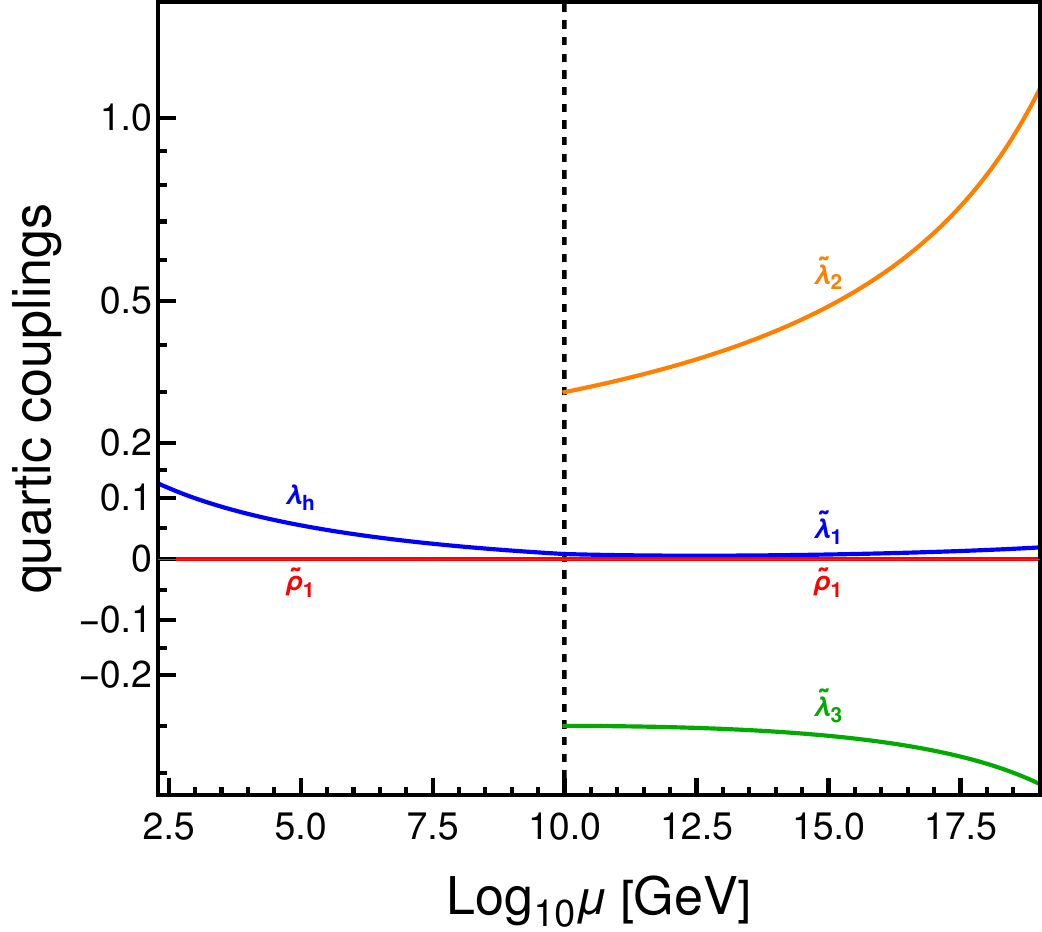}
    \caption{Evolution of the quartic couplings to show that the electroweak vacuum is stable all the way up to Planck scale. We have included full two-loop RGE effects in obtaining the results. The vertical dashed line corresponds to the mass of $H$, i.e.~$m_H=\unit[10^{10}]{GeV}$, and the LR scale $v_R=\unit[10^{18}]{GeV}$ here.}
    \label{fig:ewcure}
\end{figure}

Next we show that keeping the scalar fields below $v_R$ can cure the \textit{electroweak} vacuum stability problem. For simplicity we keep $m_{H} < v_R$. The interaction of the two doublet scalar fields $\{h, H\}$ and the singlet field $\chi$ read as
\begin{align}
   V =& -m_h^2 |h|^2 + m_\chi^2 \chi^2 + m_H^2 |H|^2 + \left\{ m_{hH}^2 h^\dagger H + \hc \right\}+ \mu_\chi \chi^3 + \mu_{H \chi} \chi |H|^2 + \mu_{h\chi} \chi |h|^2 +  \tilde{\lambda}_1 |h|^4 +  \tilde{\lambda}_2 |H|^4    \notag \\
   &+ \tilde{\lambda}_3 |h|^2 |H|^2 + \left\{ \tilde{\lambda}_4 |h|^2 H^\dagger h + \tilde{\lambda}_5 |H|^2 h^\dagger H + \tilde{\lambda}_6 h^\dagger H h^\dagger H + \hc \right\} + \tilde{\lambda}_7 h^\dagger H H^\dagger h +  \tilde{\rho}_1 \chi^4  + \tilde{\alpha}_{1} \chi^2 |h|^2 + \tilde{\alpha}_2 \chi^2 |H|^2  \, . 
\end{align}
The RGE of the quartic couplings can be written as
\begin{equation}
     \mu \frac{dX}{d\mu} \equiv \frac{1}{(4\pi)^2} \beta^{(1)} (X) \, ,
\end{equation}
where $\beta^{(1)}(X)$ are the one-loop functions of the quartic couplings which,  in the momentum range $m_H \leq \mu \leq v_R$, are given by 
\begin{align}
\beta^{(1)}(\tilde{\lambda}_1) =&

 24 \tilde{\lambda}_1^{2}

+ 2 \tilde{\lambda}_3^{2}

+ 2 \tilde{\lambda}_3 \tilde{\lambda}_7

+ \tilde{\lambda}_7^{2}

+ 2 \tilde{\alpha}_1^{2}

-  \frac{9}{5} \tilde{\lambda}_1 g_1^{2}

- 9 \tilde{\lambda}_1 g_2^{2}

+ \frac{27}{200} g_1^{4}

+ \frac{9}{20} g_1^{2} g_2^{2}

+ \frac{9}{8} g_2^{4}

+ 12 \tilde{\lambda}_1 \left|{y_t}\right|^{2}

- 6 \left|{y_t}\right|^{4} \, , \\
%%%
\beta^{(1)}(\tilde{\lambda}_2) =&

 24 \tilde{\lambda}_2^{2}

+ 2 \tilde{\lambda}_3^{2}

+ 2 \tilde{\lambda}_3 \tilde{\lambda}_7

+ \tilde{\lambda}_7^{2}

+ 2 \tilde{\alpha}_2^{2}

-  \frac{9}{5} \tilde{\lambda}_2 g_1^{2}

- 9 \tilde{\lambda}_2 g_2^{2}

+ \frac{27}{200} g_1^{4}

+ \frac{9}{20} g_1^{2} g_2^{2}

+ \frac{9}{8} g_2^{4}
 \, , \\
%%%%%%
\beta^{(1)}(\tilde{\lambda}_3) =&

 12 \tilde{\lambda}_1 \tilde{\lambda}_3

+ 4 \tilde{\lambda}_1 \tilde{\lambda}_7

+ 12 \tilde{\lambda}_2 \tilde{\lambda}_3

+ 4 \tilde{\lambda}_2 \tilde{\lambda}_7

+ 4 \tilde{\lambda}_3^{2}

+ 2 \tilde{\lambda}_7^{2}

+ 4 \tilde{\alpha}_1 \tilde{\alpha}_2

-  \frac{9}{5} \tilde{\lambda}_3 g_1^{2}

- 9 \tilde{\lambda}_3 g_2^{2}

+ \frac{27}{100} g_1^{4}

-  \frac{9}{10} g_1^{2} g_2^{2}
\notag \\
&
+ \frac{9}{4} g_2^{4}

+ 6 \tilde{\lambda}_3 \left|{y_t}\right|^{2}
 \, , \\
%%%%%
\beta^{(1)}(\tilde{\lambda}_4) =&

 24 \tilde{\lambda}_1 \tilde{\lambda}_4

+ 6 \tilde{\lambda}_3 \tilde{\lambda}_4

+ 6 \tilde{\lambda}_3 \tilde{\lambda}_5

+ 20 \tilde{\lambda}_4 \tilde{\lambda}_6

+ 8 \tilde{\lambda}_4 \tilde{\lambda}_7

+ 4 \tilde{\lambda}_5 \tilde{\lambda}_6

+ 4 \tilde{\lambda}_5 \tilde{\lambda}_7

-  \frac{9}{5} \tilde{\lambda}_4 g_1^{2}

- 9 \tilde{\lambda}_4 g_2^{2}

+ 9 \tilde{\lambda}_4 \left|{y_t}\right|^{2}

\, , \\
%%%%%
\beta^{(1)}(\tilde{\lambda}_5) =&

 24 \tilde{\lambda}_2 \tilde{\lambda}_5

+ 6 \tilde{\lambda}_3 \tilde{\lambda}_4

+ 6 \tilde{\lambda}_3 \tilde{\lambda}_5

+ 4 \tilde{\lambda}_4 \tilde{\lambda}_6

+ 4 \tilde{\lambda}_4 \tilde{\lambda}_7

+ 20 \tilde{\lambda}_5 \tilde{\lambda}_6

+ 8 \tilde{\lambda}_5 \tilde{\lambda}_7

-  \frac{9}{5} \tilde{\lambda}_5 g_1^{2}

- 9 \tilde{\lambda}_5 g_2^{2}

+ 3 \tilde{\lambda}_5 \left|{y_t}\right|^{2}
\, , \\
%%%%%
\beta^{(1)}(\tilde{\lambda}_6) =&

 4 \tilde{\lambda}_1 \tilde{\lambda}_6

+ 4 \tilde{\lambda}_2 \tilde{\lambda}_6

+ 8 \tilde{\lambda}_3 \tilde{\lambda}_6

+ 5 \tilde{\lambda}_4^{2}

+ 2 \tilde{\lambda}_4 \tilde{\lambda}_5

+ 5 \tilde{\lambda}_5^{2}

+ 12 \tilde{\lambda}_6 \tilde{\lambda}_7

-  \frac{9}{5} \tilde{\lambda}_6 g_1^{2}

- 9 \tilde{\lambda}_6 g_2^{2}

+ 6 \tilde{\lambda}_6 \left|{y_t}\right|^{2} , \\
%%%%
\beta^{(1)}(\tilde{\lambda}_7) =&

 4 \tilde{\lambda}_1 \tilde{\lambda}_7

+ 4 \tilde{\lambda}_2 \tilde{\lambda}_7

+ 8 \tilde{\lambda}_3 \tilde{\lambda}_7

+ 4 \tilde{\lambda}_7^{2}

-  \frac{9}{5} \tilde{\lambda}_7 g_1^{2}

- 9 \tilde{\lambda}_7 g_2^{2}

+ \frac{9}{5} g_1^{2} g_2^{2}

+ 6 \tilde{\lambda}_7 \left|{y_t}\right|^{2} , \\
%%%%
\beta^{(1)}(\tilde{\rho}_1) =&

 72 \tilde{\rho}_1^{2}

+ 2 \tilde{\alpha}_1^{2}

+ 2 \tilde{\alpha}_2^{2} \, , \\
%%%%
\beta^{(1)}(\tilde{\alpha}_1) =&

 12 \tilde{\lambda}_1 \tilde{\alpha}_1

+ 4 \tilde{\lambda}_3 \tilde{\alpha}_2

+ 2 \tilde{\lambda}_7 \tilde{\alpha}_2

+ 24 \tilde{\alpha}_1 \tilde{\rho}_1

+ 8 \tilde{\alpha}_1^{2}

-  \frac{9}{10} \tilde{\alpha}_1 g_1^{2}

-  \frac{9}{2} \tilde{\alpha}_1 g_2^{2}

+ 6 \tilde{\alpha}_1 \left|{y_t}\right|^{2}
\, , \\
%%%%%%
\beta^{(1)}(\tilde{\alpha}_2) =&

 12 \tilde{\lambda}_2 \tilde{\alpha}_2

+ 4 \tilde{\lambda}_3 \tilde{\alpha}_1

+ 2 \tilde{\lambda}_7 \tilde{\alpha}_1

+ 24 \tilde{\alpha}_2 \tilde{\rho}_1

+ 8 \tilde{\alpha}_2^{2}

-  \frac{9}{10} \tilde{\alpha}_2 g_1^{2}

-  \frac{9}{2} \tilde{\alpha}_2 g_2^{2} \, .
\end{align}
Here we kept only the top quark contribution arising from the SM-like Higgs $h$. Below the scale $m_H$, only the quartic couplings $\{ \tilde{\lambda}_1 \equiv \lambda_h, \tilde{\rho_1}, \tilde{\alpha}_1 \}$ contribute. Numerically, we solve the RGEs up to two-loop beta functions and show that the inclusion of the scalar field $H$ below the LR scale stabilizes the electroweak vacuum all the way up to the Planck scale, as illustrated in Fig.~\ref{fig:ewcure}. This also preserves the necessary conditions for the boundedness of the potential. The two-loop beta functions were generated with {\tt PyR@TE} package~\cite{Sartore:2020gou} and were cross-checked with the known results~\cite{Rothstein:1990qx, Maiezza:2016ybz}.
The following input values at $m_t(m_t) = 162.8$ GeV were used for the gauge couplings~\cite{ParticleDataGroup:2024cfk}:
\begin{equation}
    \alpha_1 (m_t) = 0.01024, \hspace{5mm} \alpha_2 (m_t) = 0.0340, \hspace{5mm} \alpha_3 (m_t) = 0.1094 \, . 
\end{equation}
Here $\alpha_i = g_i^2/4\pi$. The input values for the quartic couplings at the scale $\mu = m_H$ were chosen as 
\begin{align}
    \tilde{\lambda}_1 &= 7.8 \times 10^{-3} , \hspace{5mm} \tilde{\lambda}_2 = 3.0 \times 10^{-1} , \hspace{5mm} \tilde{\lambda}_3 = -3.0 \times 10^{-1}, \hspace{5mm} \tilde{\lambda}_4 = -1.8 \times 10^{-6},   \notag \\
   \tilde{\lambda}_5 &= 1.8 \times 10^{-6} , \hspace{5mm}  \tilde{\lambda}_6 = \tilde{\lambda}_7 = 0 , \hspace{5mm} \tilde{\rho}_1 =  1 \times 10^{-46}, \hspace{5mm} 
    \tilde{\alpha}_1 = 0, \hspace{5mm} 
    \tilde{\alpha}_2 =  2 \times 10^{-18} .  
\end{align}
The boundedness conditions are similar to those given in Eq.~\eqref{eq:bound}. Note that in Fig.~\ref{fig:ewcure}, although $\tilde{\lambda}_3$ is negative above $m_H$ scale, $\tilde{\lambda}_1+2\tilde{\lambda}_2+\tilde{\lambda}_3$ remains positive all the way to the $v_R$ scale, thus preserving the vacuum stability. 
The quartic couplings that are not shown in Fig.~\ref{fig:ewcure} barely evolve and play insignificant role in the running of beta function $\beta(\tilde{\lambda}_1)$.

\change{We note that above the scale \( v_R \), the relation \( \rho_1 = 4 \tilde{\rho}_1 \) holds, while the remaining quartic couplings \( \rho_{2,3,4} \) are unconstrained and can take any values within the theoretical limits derived in Section~\ref{sec:App2}. For definiteness, we choose them to be of \( \mathcal{O}(1) \), ensuring that the scalar fields \( \delta_L^{0r} \), \( \delta_L^{0i} \), \( \delta_L^{+} \), \( \delta_L^{++} \), and \( \delta_R^{++} \) acquire masses at or above the \( v_R \) scale. See Eqs.~\eqref{eq:Neutmat}--\eqref{eq:massdouble} for the corresponding analytical expressions.}

%\end{widetext}
\twocolumngrid
%%%%%%%%%%%%%%%%%%%%%%%%%
%%%%%%%%%%%%%%%%%%%%%%%%%
\bibliographystyle{utphys}
\bibliography{BIB.bib}

\end{document}